\documentstyle[prb,aps,psfig]{revtex}
\newlength{\figwidth}
\newlength{\refspace}
\def\eps{\epsilon}
\def\half{\case{1}{2}}
\def\eb{\epsilon\beta_1}
\def\arcsinh{{\rm arcsinh}}
\begin{document}
\flushbottom
\twocolumn[
\hsize\textwidth\columnwidth\hsize\csname @twocolumnfalse\endcsname

\title{Flux-Induced Vortex in Mesoscopic Superconducting Loops}
\author{Jorge Berger$^a$ and  Jacob Rubinstein$^b$ }
\address{$^a$Department of Physics, Technion, 32000 Haifa, Israel\\
$^b$Department of Mathematics, Technion, 32000 Haifa, Israel}
\maketitle{
\begin{abstract}
We predict the existence of a quantum vortex for an unusual situation.
We study the order parameter in doubly connected superconducting samples
embedded in a uniform magnetic field. For samples with 
perfect cylindrical symmetry, the order parameter has been
known for long and no vortices are present in the linear regime. However,
if the sample is not symmetric, there exist ranges of the
field for which the order parameter vanishes along a line, parallel to the
field. In many respects, the behavior of this line is qualitatively different
from that of the vortices encountered in type II superconductivity.
For samples with mirror symmetry, this flux-induced vortex appears at
the thin side for small fluxes and at the opposite side for large fluxes. 
We propose direct and indirect experimental methods which could test our
predictions.
\end{abstract}}{
\pacs{}}  ]                     
\vspace{-0.2cm}
The question of where and under what conditions an individual vortex appears
(or disappears) is still an active subject
\cite{hist,Parks,Bean,num,rev,he}.
Here we consider an unusual setup: that of the Little-Parks experiment
\cite{LP,rings,Price}, {\it i.e.} a superconducting loop which encloses a
magnetic flux.

We find that, for a narrow range of enclosed fluxes in the vicinity of a
half-integer number of quanta, a vortex is present in the loop.
This vortex is completely different in
nature from the traditional vortices known from type II superconductivity.
In the traditional case, a minimum local magnetic field is required for the
formation of a vortex; here, its formation is governed by the enclosed flux.
As the magnetic field is increased, instead of having more and more vortices
coming in, the present vortex disappears (and a new vortex appears only after
an additional quantum of flux is enclosed by the loop). Whereas traditional
vortices cannot exist along films which are thinner than the coherence length
(they ``need room for their cores") \cite{Parks}, the present vortex exists
for arbitrarily thin shells. Another atypical feature of the present 
vortex is its high anisotropy, which depends on the thickness of the shell.
Moreover, there are cases in which the order parameter vanishes {\em on a
surface} rather than a line.

Additional differences arise from the assumption that the total magnetic field 
remains equal to the applied field: whereas traditional vortices exist for 
type II 
superconductors, for which the Ginzburg-Landau parameter $\kappa$ is larger
than $2^{-1/2}$, the present vortex is insensitive to the London penetration 
depth and bears no connection with $\kappa$; whereas traditional vortices 
cannot stay in 
equilibrium if the distance to the boundary is too short \cite{Bean}, the
equilibrium position of the present vortex is a continuous function of the
flux and does reach the boundaries. This assumption is justified when either
the thickness or the width of the loop is small compared to the magnetic
penetration depth. The case in which the induced field is not negligible will
be considered elsewhere.

We shall consider two complementary geometries: samples which almost have
axial symmetry, which will be analyzed using perturbation, and samples with
eccentric cylindrical boundaries, which will be studied by means of variation.
In the first method the zeroth order configuration is a cylindrical shell of 
superconducting material, with inner
radius $R$, outer radius $R'=R+w$ and heigth $h$, where $R$, $w$ and $h$ are
constants. In early experiments \cite{LP} $h$ was much larger than $R$ and 
$w$, whereas recent experiments \cite{rings,Price} meet the opposite 
situation. We consider a uniform magnetic field parallel to the axis of the
shell and study the onset of superconductivity. In this regime the magnetic
field is still unaffected by the supercurrents and the Ginzburg-Landau 
equation is linearized to the eigenvalue problem
\begin{equation}
H_0 \psi_0=\left(i\nabla-2\pi{\bf A}/\Phi_0 \right)^2 \psi_0=
\psi_0/\xi_0^2=E_0 \psi_0 .
\label{Sch}
\end{equation}
$\psi_0$ is the order parameter, ${\bf A}$ is the magnetic potential, $\Phi_0$ 
is the quantum of fluxoid ($\Phi_0>0$) and $\xi_0$ is the coherence length at
the onset of superconductivity.
The notations for the function $\psi_0$, the operator $H_0$ and the 
eigenvalue $E_0$ have been introduced to remind of the eigenfunction,
the Hamiltonian and the energy in a perturbation problem.
Eq.~(\ref{Sch}) is 
subject to the condition that $(i\nabla-2\pi{\bf A}/\Phi_0)\psi$ has no 
normal component at the boundaries. (For the moment, $\psi=\psi_0$.) 
For a cylindrical shell and cylindrical coordinates $(r,\theta,z)$,
Eq.~(\ref{Sch}) is separable and has been solved in [\onlinecite{cyl}]. 
Writing $\psi_0={\cal R}(r)\Theta (\theta){\cal Z}(z)$, one easily obtains
${\cal Z}(z)=\cos(k\pi z/h)$ and $\Theta (\theta)=e^{-mi\theta}$,
with $k$ and $m$ integers. Since $E_0$ always increases with $k^2$, we
pick $k=0$ and the solution of Eq.~(\ref{Sch}) reduces to 
\begin{equation}
\psi_0={\cal R}_m(r) e^{-mi\theta}.
\label{tet-r}
\end{equation}
The winding number $m$ is chosen so that lowest value of $E_0$ is
obtained. For a thin shell it is the closest integer to the number of flux
quanta enclosed by the shell.
${\cal R}_m$ was obtained in [\onlinecite{cyl}]; it depends on the 
magnetic field and involves a combination of Kummer functions, adjusted to
meet the boundary conditions. 

A salient feature of the solution (\ref{tet-r}) is that there exist
magnetic fields for which Eq.~(\ref{Sch}) is degenerate. In the limit of a
thin shell this occurs when the enclosed flux $\Phi$ equals
$(m+\half)\Phi_0$: for this flux, the eigenvalues $E_0$ obtained for $m$
and for $m+1$ are the same and the lowest possible. If the shell is not
thin, we can still define $\Phi$ as the magnetic flux enclosed
by the ``representative circle" $r=\bar R=R+w/2$. There still exist fluxes
$\Phi^*_m$ (usually close to $(m+\half)\Phi_0$) where the
values of $E_0$ for $\psi_0={\cal R}_m(r)\exp(-mi\theta)$ and for
$\psi_0={\cal R}_{m+1}(r)\exp(-[m+1]i\theta)$ coalesce.
As the magnetic field is swept accross the situation $\Phi=\Phi^*_m$, the
order parameter $\psi_0$ changes discontinuously. With it, there is a jump
in measurable quantities such as the current around the shell.  

We perturb now the problem and consider a sample which is not a perfect 
cylindrical shell. To fix ideas, we still keep $h$ and $R$ constant, but
take a nonuniform width
\begin{equation}
D(\theta)=w\left(1+\frac{\eps}{2}\sum_{j\ne 0}\beta_j e^{ji\theta}\right),
\label{D}
\end{equation}
where $w$ is the average width, $\eps \ll 1$ and $\beta_{-j}=\bar\beta_j$.
Other deviations from the perfect cylindrical problem lead to similar
behavior.

The ``Hamiltonian" is still $H_0$, but the eigenvalue and the eigenfunction
are perturbed by the change in the boundary geometry. We write
$\psi=\psi_0+\eps\psi_1+\eps^2\psi_2+\ldots$ and 
$E=E_0+\eps E_1+\eps^2E_2+\ldots$, where $\psi$ is the order parameter and
$E^{-1/2}$ the coherence length. The eigenvalue in our problem is
related to the temperature $T$ by
\begin{equation}
R^2 E=(T_c-T)/(T_c-T_R),
\label{temp}
\end{equation}
where $T_c$ is the critical temperature in the absence of magnetic field
and $T_R$ the temperature at which the
coherence length equals the internal radius $R$.
The only reason for the requirement of a ``mesoscopic" sample is to
decrease $T_R$ and thus widen
the temperature scale $T_c-T_R$; our formalism is valid for arbitrary
positive $R$, $w$ and $h$, and may cover the
entire range from a very thin ring to an almost full disk. 
In order to decrease $T_R$ we also require clean materials (usually type
I), which have a large coherence length at zero temperature.

To proceed, we define a metric by integrating in the {\em unperturbed} 
region: 
$(\phi_1,\phi_2)=\int_0^{2\pi}\!d\theta\!\int_R^{R'}rdr\bar\phi_1\phi_2$.
It follows that
\begin{equation}
(\phi_1,H_0\phi_2)-(H_0\phi_1,\phi_2)=R'\int_0^{2\pi} \left(\phi_2
\frac{\partial\bar\phi_1}{\partial r}-\bar\phi_1 \frac{\partial\phi_2}
{\partial r}\right)d\theta.
\label{Herm}
\end{equation}

Far from $\Phi_m^*$, we obtain the sequence of equations
$(H_0-E_0)\psi_1=E_1\psi_0$,
$(H_0-E_0)\psi_2=E_1\psi_1+E_2\psi_0$, etc. The difference between this 
procedure and standard perturbation theory is that $H_0$ is {\em not 
Hermitian}; instead, it obeys Eq.~(\ref{Herm}). To take this into account,
we need $\partial\psi_i/\partial r$, evaluated at $r=R'$. To obtain this we
expand $\psi$ around $r=R'$ and require to all orders of $\eps$ that the
normal current vanishes at $r=R+D(\theta)$. In this way we obtain that 
$E_1=0$; the expressions for $\psi_1$ and $E_2$ are lengthy and will be 
reported elsewhere.

For $\Phi\approx\Phi_m^*$, the direct perturbation scheme described in the
previous paragraph would lead to a divergent $\psi_1$. This divergence is
due to the degeneracy at $\Phi=\Phi_m^*$. In this case we use
degenerate perturbation theory, {\it i.e.} we write
\begin{equation}
\psi_0={\cal R}_m(r)\exp(-mi\theta)-
\gamma {\cal R}_{m+1}(r)\exp(-[m+1]i\theta) ,
\label{degen}
\end{equation}
with the normalization ${\cal R}_m(R)={\cal R}_{m+1}(R)=1$ and $\gamma$ a
coefficient which still has to be determined. Writing
$2\pi R^2{\bf A}=\Phi_0[b_m+\eps\delta]r\hat\theta$, with
$\bar R^2b_m=\Phi^*_m/\Phi_0$, and substituting into
 (\ref{Herm}) $\phi_1$ by $\psi_0$ and $\phi_2$ by ${\cal
R}_m(r)\exp(-mi\theta)$ or ${\cal R}_{m+1}(r)\exp(-[m+1]i\theta)$,
leads to a system of equations for $\gamma$ and $E_1$:
\begin{eqnarray}
A_m\delta+B_m E_1&=&C_m \beta_1 \gamma \ , \\
A'_{m+1}\delta+B_{m+1} E_1&=&C_m \bar\beta_1/\gamma \ ,
\label{system}
\end{eqnarray}
where
\begin{eqnarray}
A_m&=&2\int_R^{R'}(mr-b_m r^3){\cal R}_m^2 dr \ ,  \\
A'_m&=&2\int_R^{R'}(mr-b_{m-1} r^3){\cal R}_m^2 dr \ , \\
B_m&=&\int_R^{R'}r{\cal R}_m^2 dr \ , \\
C_m&=&\half wR'{\cal R}_m(R'){\cal R}_{m+1}(R') \times \nonumber \\
&&[E_0-(b_mR'-(m+1)/R')(b_mR'-m/R')]
\end{eqnarray}
and $R$ has been taken as the unit of length. This system has two
solutions and we choose the one with lower $E_1$. Note that this time
$E_1\neq 0$. After obtaining $\psi_0$ and $E_1$, $\psi_1$ and $E_2$ were
also evaluated, but the details will be reported elsewhere.

\begin{figure}[bth]
\centerline{\psfig{figure=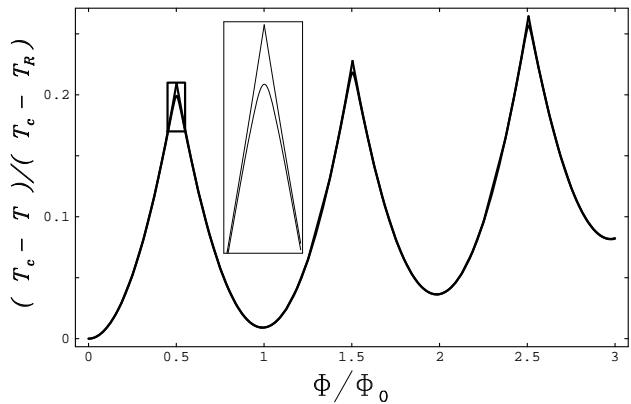,width=\figwidth}}
\vspace*{0.2cm}
\caption{
Temperature $T$ for the onset of superconductivity as a function of the
magnetic flux $\Phi$ enclosed by the ``representative" circle. The graph
shows two curves, which nearly coincide. The lower curve
corresponds to the formalism developed here; for comparison, 
$E_{\rm const}$ (see text) has also been drawn. 
The inset shows an enlarged view of the region enclosed by the rectangle.
The lower curve in the inset looks smooth, but was calculated using three
different algorithms in different regions: $m=0$, degenerate perturbation
and $m=1$; likewise, near $\Phi=\Phi_1^*\approx 1.5\Phi_0$, $m=1$,
degenerate perturbation and $m=2$ match smoothly, and so on.}
\label{onset}
\end{figure}

The value of $E$ provides the temperature at which the shell becomes 
superconducting. We have calculated it for a sample with a reasonable 
experimental shape: $D(\theta)=0.2R-0.01R\cos(\theta)$. The results are 
shown in Fig.~\ref{onset}. A customary approximation ({\it e.g.} 
[\onlinecite{rings}])
for results in a thick shell is obtained by averaging the one-dimensional
result for ideally thin shells. This procedure gives
$E_{\rm av}=\int(\pi Br/\Phi_0-m/r)^2dr/w$, where $B$ is the magnetic field,
$m$ is the integer that minimizes this expression and the integral is from
$R$ to $R'$. 
A somewhat better approximation is obtained by taking 
${\cal R}_m$ constant ({\it e.g.} [\onlinecite{Price}]). This gives $E_{\rm
const}=\int(\pi Br/\Phi_0-m/r)^2rdr/\!\int rdr$.  
$E_{\rm const}$ is also shown in Fig.~\ref{onset} for comparison.
We see that $E_{\rm const}$ is quite a good
approximation, but it exhibits cusps when the value of $m$ changes, whereas
for our treatment $E$ is smooth. This qualitative difference does not arise
from the dependence of ${\cal R}_m$ on $r$; what happens is that the 
superposition of two values of $m$ produces a sinusoidal dependence of 
$|\psi|^2$ on $\theta$.\cite{SIAM} The size and the extent of the
decline of $T_c-T$ near $\Phi_m^*$ are proportional to $\eb$, which
measures the deviation from a uniform shell.

\begin{figure}[bth]
\centerline{\psfig{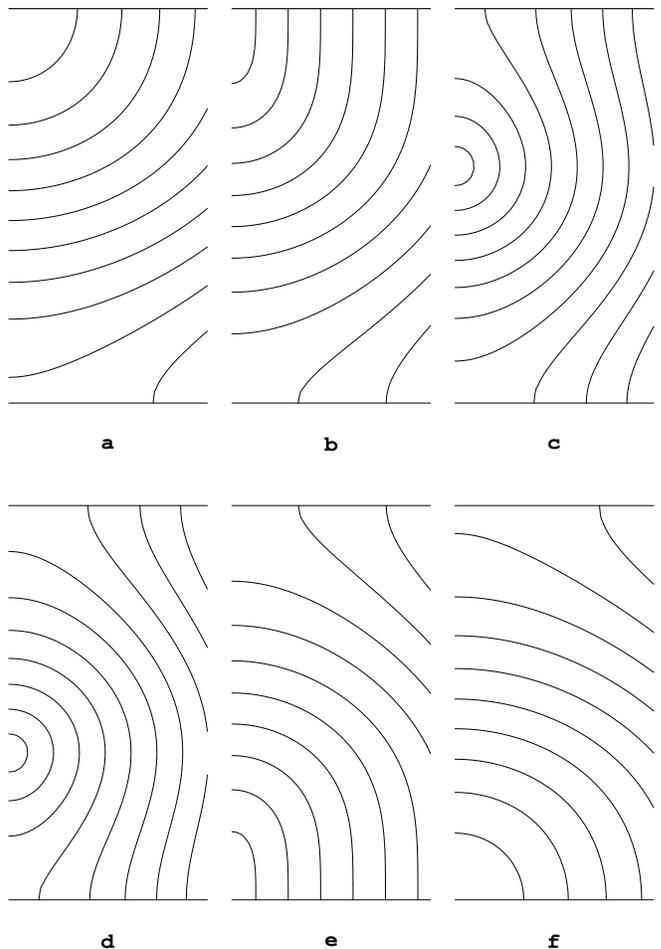}}
\vspace*{0.2cm}
\caption{
Contour plot of the absolute value of the order parameter in a narrow
radial slice of the superconducting shell, for increasing values of the
magnetic field. The horizontal coordinate is $|\theta|$, in the region
$0\le|\theta|\le 0.0002\pi$; the vertical coordinate is $R\le r \le
1.19R$. The plane $\theta=0$ is located at the left of each frame. In (a),
$|\psi|$ has a minimum at the line $(r=1.19R,\theta=0)$. In (b), $\psi$
vanishes along this line in the outer boundary and the vortex appears. As
the magnetic field increases, the vortex shifts towards the inner boundary 
$r=R$ and reaches it in (e). In (f), $|\psi|>0$ everywhere and the vortex
has disappeared.}
\label{psi}
\end{figure}

In Fig.~\ref{psi} we show contour plots of $|\psi|$ in a narrow radial strip 
next to $\theta=0$ for six equidistant fluxes between 
$\Phi=0.501364 \Phi_0$ and $\Phi=0.501382 \Phi_0$. 
$\Phi_0^*=0.501376\Phi_0$ is contained in this range.
We have chosen an example with the width $D$ symmetric about
$\theta=0$ and $\theta=\pi$, so that the plots can be continued by placing
a mirror at $\theta=0$. As $\Phi$ increases and comes close to $\Phi_0^*$,
$|\psi|$
decreases at the axial line $(r=R+D(0),\theta=0)$, until in frame (b)
$\psi(R+D(0),0)=0$: this is the field at which the vortex appears. As the
flux is increased further, the equilibrium position of the vortex moves
towards the inner boundary, which is reached in frame (e). If the field is
increased further, the vortex disappears.
The extent of the flux range for which the vortex exists is
proportional to $\eb$.

In the general case the vortex is located at
$\theta=\arg(-C_m\bar\beta_1)$;
if $m$ is small and $D$ is symmetric, as in the example we picked, this
corresponds to the thinnest part of the shell.
For large values of $m$ ($m\ge 5$ for $w=0.2R$),
$C_m$ becomes negative and the vortex is located at the opposite side of
the shell. 

\begin{figure}[bth]
\centerline{\psfig{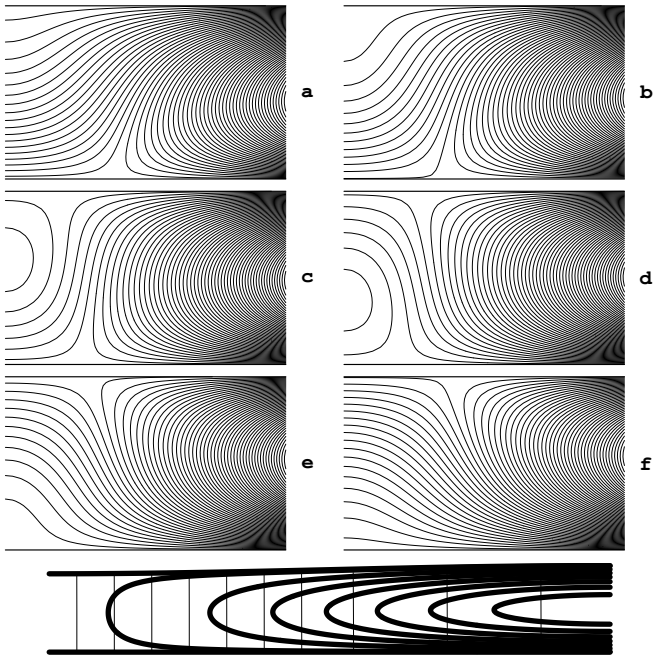}}
\vspace*{0.3cm}
\caption{
Streamlines in a piece of the superconducting shell, for the same magnetic
fields as in Fig.~\protect{\ref{psi}}. The coordinates are the same as in
Fig.~\protect{\ref{psi}}, but now $0\le|\theta|\le 0.1\pi$. In (a) and (b)
we see streamlines through $\theta=0$ that carry electric current to the
right (assuming that the magnetic field comes out of the page). These
streamlines circulate around the superconducting loop (enclose the hollow
region). At the right we see part of the screening currents, in clockwise
circuits that do not enclose the hollow region. In (c) we see two 
counterclockwise streamlines around the quantum vortex, three streamlines
around the loop and the screening currents at the right.
Along a path around the vortex the phase of the order parameter changes by
$2\pi$. In (d) the vortex
has moved towards the inner boundary; the current around the loop has now
become counterclockwise. As the magnetic field increases, the vortex moves
down until it disappears and the counterclockwise current around the loop
increases. The bottom frame shows a ``panoramic" view of half of the
shell, $0\le|\theta|\le\pi$. The magnetic field is the same as in (d). The
thick lines are streamlines and the thin lines are lines of constant
$|\psi|$. On this scale, $|\psi|$ looks independent of $r$ and the
currents around the vortex and around the loop do not show up.}
\label{stream}
\end{figure}

Knowing $\psi(r,\theta)$, we can evaluate the supercurrent. Fig.~\ref{stream}
shows the streamlines for the same fluxes considered in Fig.~\ref{psi}. Each
frame shows 1/20 of the shell, with the plane $\theta=0$ at the left.

\begin{figure}[bth]
\centerline{\psfig{figure=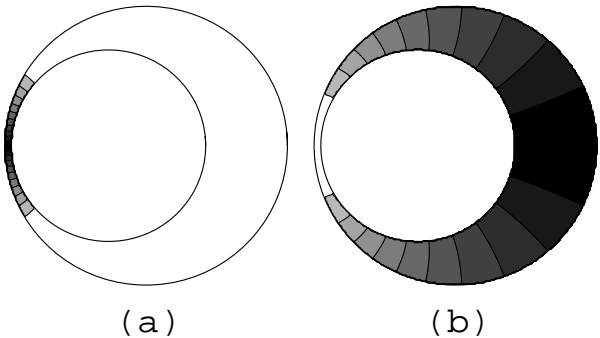,width=\figwidth}}
\vspace*{0.3cm}
\caption{
Contour plot of $|\psi|$ for a sample with cylindrical inner and outer 
boundary. Darker areas denote smaller values of $|\psi|$. For the chosen
flux values, $\psi=0$ roughly in the middle of the darkest areas. 
(a) $\Phi=2.5713\Phi_0$. (b) $\Phi=3.61\Phi_0$. The flux range for the
presence of the vortex in case (b) is about forty times larger than in 
case (a).}
\label{bipolar}
\end{figure}

The perturbational approach is theoretically instructive, but since the range
of existence of the vortices increases with nonuniformity, we are interested
in sample shapes that are very far from axial symmetry. We conjecture that
the scenario found above is generic for samples that are not symmetric. To
support this conjecture, we consider a sample with cylindrical boundaries.
The inner (outer) radius will again be denoted by $R$ ($R'$), but the axes of 
both boundaries do not coalesce. $\Phi$ will still be the flux through a
circle of radius $\bar R=\half(R+R')$. It is natural to use bipolar 
coordinates,\cite{bip} $x+iy=c\tanh[(\alpha+i\beta)/2]$, where $x$ and $y$
are the Cartesian coordinates perpendicular to the axes and $c$ a constant
which defines the lengthscale. The lines of constant $\alpha$ are circles;
we assign $\alpha_1=\arcsinh(c/R')$, $\alpha_2=\arcsinh(c/R)$, and the sample
occupies the region $\alpha_1\le\alpha\le\alpha_2$, $-\pi\le\beta\le\pi$.
Thinner
samples (smaller $(R'-R)/R$) are obtained by taking small differences 
$\alpha_2-\alpha_1$, whereas more eccentric samples (larger $D(\pi)/D(0)$)
are obtained by taking both $\alpha_1$ and $\alpha_2$ small. The shape
shown
in Fig.~\ref{bipolar} corresponds to $\alpha_1=0.5$, $\alpha_2=0.7$.

The magnetic potential can be written in the form
\begin{eqnarray}
{\bf A}&=&\frac{\Phi (\cosh\alpha+\cos\beta)}{\pi c\bar R^2}\left[f(\alpha,\beta)
-f(\bar\alpha,\beta)-\frac{\bar R^2}{2}\right]{\bf \hat\beta}, \nonumber
\\
f(\alpha,\beta)&=&c^2\sin^{-2}\!\beta\;[\sinh\alpha/(\cosh\alpha+\cos\beta)
\nonumber \\ &-&
2\arctan(\tan\case{\beta}{2}\tanh\case{\alpha}{2})\cot\beta],
\label{potential}
\end{eqnarray}
with $\bar\alpha=\arcsinh(c/\bar R)$.

An approximation for the order parameter $\psi$ will be obtained by means of
a variational procedure.\cite{variations} The eigenvalue and the eigenvector
for the onset of superconductivity may be obtained by minimizing the ratio
$\int|(i\nabla-2\pi{\bf A}/\Phi_0)\psi|^2dV/\int|\psi|^2dV$, where $dV$ is
the element of volume and the integrals are over the sample. Using the
boundary conditions, $\psi$ can be written as a Fourier series
\begin{eqnarray}
\psi=&&\sum_{m=-\infty}^\infty\sum_{l=0}^\infty\left(
c_{ml}\cos\left(l\pi\frac{2\alpha-\alpha_2-\alpha_1}{\alpha_2-\alpha_1}
\right)\right.\nonumber \\ &+&\left.
s_{ml}\sin\left(\frac{(2l+1)\pi}{2}\frac{2\alpha-\alpha_2-\alpha_1}
{\alpha_2-\alpha_1}\right)\right)e^{mi\beta}.
\label{Fourier}
\end{eqnarray}
($\psi$ is again independent of $z$.) We truncate now this series and
leave
only a finite set of nonzero coefficients. The number of terms which is 
required to achieve some desired accuracy increases with $\Phi$ and with
$(R'-R)/R$, while it is not very sensitive to eccentricity. In our
calculations we have kept eight coefficients: $c_{m_0-2,0}$, $c_{m_0-1,0}$, 
$c_{m_0,0}$, $c_{m_0+1,0}$, $c_{m_0+2,0}$, $s_{m_0-1,0}$, $s_{m_0,0}$, 
$s_{m_0+1,0}$, where the integer $m_0$ is chosen to obtain the minimum
eigenvalue.

Regarding these coefficients as a vector $${\bf v}=\{c_{m_0-2,0}, 
c_{m_0-1,0}, c_{m_0,0}, c_{m_0+1,0},$$ $$c_{m_0+2,0}, s_{m_0-1,0},
s_{m_0,0}, 
s_{m_0+1,0}\},$$ we
can write $\int|(i\nabla-2\pi{\bf A}/\Phi_0)\psi|^2dV={\bf v^\dagger Mv}$ and
$\int|\psi|^2dV={\bf v^\dagger Qv}$, where the elements of the matrices 
${\bf M}$ and ${\bf Q}$ are easily evaluated. We are therefore left with
the minimization of the ratio $({\bf v^\dagger Mv)/(v^\dagger Qv})$. This
ratio is minimized for
the eigenvector of ${\bf Q}^{-1}{\bf M}$ that has the lowest eigenvalue.
The error introduced by truncation may be estimated by the size of the
discontinuity of $\psi$ at the values of $\Phi$ for which $m_0$ changes.

Using this method, we rediscover the behavior obtained for almost axially
symmetric samples: near half-integer fluxes, a vortex appears at the
outer
boundary and moves towards the inner boundary as the flux is increased.
What
looks more striking, is that again these vortices enter through the thin side
for small numbers of quanta and through the opposite side when $\Phi/\Phi_0$
is sufficiently large. If $\alpha_{1,2}$ are chosen to reproduce the shape
used in our previous perturbation example, the length of the range for which
the vortex is present agrees to about $10^{-8}\Phi_0$ with the result 
obtained by perturbation. The values of the flux for which the vortex appears
agree to about $10^{-6}\Phi_0$ in both methods. Fig.~\ref{bipolar} shows
contour plots of $|\psi|$ for $\alpha_1=0.5$, $\alpha_2=0.7$ and for fluxes
chosen so that the vortex is approximately midway between the inner and the
outer boundary. For these values of $\alpha_{1,2}$ and for $\Phi\le 5\Phi_0$,
we estimate the error in $\psi$ to be a few percent of the maximum value of
$|\psi|$.

In conclusion, we have found a setup for which a new kind of vortex exists.
Its existence is governed by the flux enclosed in the entire 
shell rather than by the local magnetic field.
Likewise, the position of the vortex is a function of the 
entire sample (and of the flux) rather than of local defects. 
The ocurrence of this vortex is a nearly periodic function of the flux.
The presence of a vortex implies that the flux lies within a very narrow
known range.

It should be stated, however,
that the magnetic field in the superconducting material itself does
influence
the nature of these vortices. For instance, we have found that if the
field is constant in the 
region $0\le r\le\bar R$ and has opposite sign outside this region, then,
as $\Phi$ 
increases close to $\Phi_m^*$, a vortex and an antivortex form at a point
somewhere in the middle of the shell; as the flux increases further, the
vortex moves from this point towards the inner boundary and the antivortex
towards the outer boundary, until they disappear. Moreover, if the field
vanishes in the entire shell, then the vortex becomes a
{\em cut},\cite{cut} {\it i.e.} the order parameter vanishes on a surface
that connects the inner and the outer boundaries.

Finally, we discuss the experimental possibilities for the detection of the
vortex predicted here. We require a sample such that its size is of the order 
of the coherence length for a reasonable range of temperatures. Typical
experiments for this purpose use samples made of Al, with a perimeter
of a few $\mu$m. In order to observe the vortices directly, we
require an imaging technique with spatial
resolution of the order of 0.1 $\mu$m. This requirement is met by scanning
tunneling microscopy,\cite{STM} magnetic decoration\cite{decor} and 
electron holography.\cite{holo}
When interpreting
electron holography it should be born in mind that for the vortex predicted
here only the fluxoid equals $\Phi_0$; its magnetic flux alone will be much
smaller and temperature dependent.
Clearly, in order to sense the local density of states or the current 
density, the size of the order parameter must be above some threshold 
required by the sensitivity of the experiment. In this regime
Eq.~(\ref{Sch}) for the onset of superconductivity becomes insufficient and 
we have to analyze the nonlinear Ginzburg-Landau equations. We have partially
analyzed this situation and anticipate here some results. 
The existence and
behavior of the vortex remain qualitatively the same as reported here for
some temperature range below the onset of superconductivity. This temperature 
range increases with the deviation from a uniform shell. 

Direct observation of the vortex is expected to be a difficult experiment,
since the requirement of large coherence lengths implies a small gap. In 
addition, evaporated Al usually has a rough surface. Therefore, it might be
helpful to obtain indirect evidence for the scenario encountered here.
In the limit of thin shells, we know that there exist critical points at
fluxes close to $\Phi_m^*$ and at some temperature which we denote 
$T_2^{(m)}$. Below $T_2^{(m)}$ the current $I$ around the loop exhibits
hysteresis, whereas above $T_2^{(m)}$ it is reversible and continuous.
Moreover, above and near $T_2^{(m)}$, $dI/d\Phi$ has a maximum near 
$\Phi=\Phi_m^*$, which diverges as $T_2^{(m)}$ is approached.\cite{PRB}
Our analysis of the nonlinear Ginzburg-Landau equations shows that these
features remain valid when the shell is thick. In particular, the divergence
of $dI/d\Phi$ at the critical point is not smeared.

We therefore need a technique that measures $dI/d\Phi$, which is proportional
to the ac magnetic susceptibility. This is measured by applying a bias flux
with a superimposed small ac signal. The response is the ac component of the
induced magnetic flux in some region. This induced flux has already been 
measured by means of a SQUID microsusceptometer.\cite{Price} Other
techniques which seem
appropriate for this measurement are the piezoresistive
cantilever\cite{vibr}
and the ballistic Hall magnetometer.\cite{HPM} The Hall magnetometer seems
particularly promising,
because, if only part of the sample is located on the active region of the
probe, it would allow to measure the flux induced at that part.

This research was supported by the US-Israel Binational Science
Foundation and by the Israel Science Foundation. We thank 
A. Auerbach, Y. Eckstein, B. Fisher and V. Kogan for useful suggestions.
We thank V. Bruyndoncx, V. Moshchalkov, B. Pannetier and C. Van
Haesendonck
for clarifying to us some of the experimental possibilities and limitations.


\end{document}